\documentclass[onecolumn,showpacs,amsmath,amssymb,subfigure, 12pt]{revtex4}
\usepackage{graphicx}
\usepackage{hhline}
\usepackage{mathrsfs}
\usepackage{amssymb}
\usepackage{setspace}
\usepackage[newitem,newenum]{paralist}
  {\pointedenum\begin{enumerate}}%
  {\end{enumerate}}
  {\pointlessenum\begin{enumerate}}%
  {\end{enumerate}} 
\usepackage{bm}
\usepackage[dvips]{color}
\usepackage{color}
\begin{document}
\hspace{-1.5em}\Large{\textbf{Thermal fluctuations of granular gas under HCS using two-point kinetic theory}}\\ \\
\large\textmd{Ryosuke Yano}\\ \\
\small\textit{Department of Advanced Energy, University of Tokyo, 5-1-5 Kashiwanoha, Kashiwa, Chiba 277-8561, Japan, Email: yano@daedalus.k.u-tokyo.ac.jp}\\ \\
\large\textmd{Kojiro Suzuki}\\ \\
\small\textit{Department of Advanced Energy, University of Tokyo, 5-1-5 Kashiwanoha, Kashiwa, Chiba 277-8561, Japan, Email: suzuki@k.u-tokyo.ac.jp}
\begin{center}
\textsf{Abstract}
\end{center}
\begin{abstract}
Thermal fluctuations of the granular gas under the homogeneous cooling state (HCS) are estimated using two-point kinetic theory by Tsuge-Sagara. Thermal fluctuations of the elastic gas are modified for the granular gas by nonequilibrium moments, which defines the distribution function under the HCS. The deviations of thermal fluctuations for the granular gas from those for the elastic gas obtained by the fluctuation-dissipation theorem are calculated as a function of the restitution coefficient.
\end{abstract}
\maketitle
\section{Introduction}
The granular gas becomes a popular issue owing to its rich physical phenomena \cite{Brilliantov}. Especially, the pattern formation via the inelastic collisions is presumably caused by the local thermal fluctuations and fluid dynamic fluctuations. The decay of the amplitude of fluctuations via the decrease of the temperature as a  result of inelastic collisions yields a larger pattern formation via the dissipation. On the other hand, properties of thermal fluctuations for the granular gas are not understood completely, whereas Brey et al. \cite{Brey} formulated the thermal fluctuations on the basis of the the linear Boltzmann equation with a fluctuating white-noise term. In this paper, we consider on deviations of thermal fluctuations for the granular gas from those for the elastic gas. To simplify our discussions, we restrict ourselves to the HCS \cite{Brey}. To evaluate thermal fluctuations for the granular gas under the HCS, the two-point kinetic theory by Tsuge-Sagara \cite{Tsuge} for the elastic gas is extended into the granular gas. Assuming the HCS, the correlation between different particles, which yields fluid dynamic fluctuations \cite{Tsuge}, can be neglected, and only self-correlation function \cite{Tsuge} is considered exclusively. Finally, deviations of thermal fluctuations for the granular gas from those for the elastic gas are obtained as a function of nonequilibrium moments, which define the distribution function under the HCS. In this letter, the two-point kinetic theory is revisited in detail, and extended into the granular gas. Finally, we evaluate deviations of thermal fluctuations for the granular gas from those for the elastic gas as a function of the restitution coefficient. In this paper, inelastic hard sphere particles with a constant restitution coefficient and the smooth surface are discussed.
\section{Revisited two-point kinetic theory by Tsuge-Sagara \cite{Tsuge}}
The two-point kinetic theory is proposed by Tsuge-Sagara, and discussed for the elastic gas. In this section, we revisit the two-point kinetic theory and introduce fluctuation-dissipation (FD) theorem \cite{Landau} for thermal fluctuations of the elastic gas. The stochastic Boltzmann equation is the most fundamental equation in the two-point kinetic theory. The stochastic Boltzmann equation is written as follows:
\begin{eqnarray}
&&B \varrho(t,\zeta)=\left(\frac{\partial}{\partial t}+\bm{c} \frac{\partial }{\partial \bm{x}}\right)\varrho(t,\zeta)-\mathscr{J}\left(\hat{\zeta}|\zeta\right)\left[ \varrho\left(t,\zeta\right)\varrho\left(t,\hat{\zeta}\right)\right]=0,\\
&&\mathscr{J}\left(\hat{\zeta}|\zeta\right)\left[\varrho\left(t,\zeta\right)\varrho\left(t,\hat{\zeta}\right)\right] \nonumber \\
&=&\int_{\mathscr{V}^3} d\hat{\bm{c}} \int_0^{2\pi} d\tilde{\epsilon} \int_0^\pi d\chi \left\{\varrho\left(t,\zeta^\prime\right)\varrho\left(t,\hat{\zeta}^\prime\right)-\varrho\left(t,\zeta\right)\varrho\left(t,\hat{\zeta}\right) \right\}g\sigma \sin \chi, \nonumber \\
\end{eqnarray}
where $\varrho(t,\zeta)$ is the microscopic distribution function defined by:
\begin{eqnarray}
\varrho(t,\zeta)=\sum_{s=1}^N \delta \left[\zeta-\zeta^{(s)}(t) \right],
\end{eqnarray}
where $s$ is the index of the partiles, $N$ is the total number of particles, $\zeta=\left(\bm{c},\bm{x}\right)$, in which $\bm{c}\in \{\mathscr{V}^3: -\bm{\infty} < \bm{c} <\bm{\infty}\}$ is the velocity space and $\bm{x}$ is the physical space. In eq. (1), the prime indicates the state after binary collisions, $g=|\bm{c}-\hat{\bm{c}}|$ is the relative velocity of the binary colliding particles, $\sigma$ is the differential collision-cross section, $\chi$ is the deflection angle, $\tilde{\epsilon}$ is the scattering angle.\\
For convenience, we set $\xi=(t,\zeta)$. In $\xi$-space, we consider the correlation between two points $"\alpha"$ and $"\beta"$. As a result, $\varrho(\alpha)\equiv\varrho(\xi(\alpha))$ and $\varrho(\beta)\equiv \varrho(\xi(\beta))$ are used. At point $\alpha$, the stochastic Boltzmann equation in eq. (1) is rewitten as:
\begin{eqnarray}
B(\alpha)\varrho\left(\alpha\right)=\left[\frac{\partial}{\partial t(\alpha)}+\bm{c}(\alpha)\frac{\partial}{\partial \bm{x}(\alpha)}\right]\varrho(\alpha)-\mathscr{J}(\hat{\alpha}|\alpha)\left[\varrho(\alpha)\varrho(\hat{\alpha})\right]=0.
\end{eqnarray}
Next, we define the averaged quantites in $\zeta$-space as follows using finite volumes $\Delta V \in \bm{x}$ and $\Delta V_c \in \mathscr{V}^3$:
\begin{eqnarray}
\overline{\psi}&=&\frac{1}{\Delta V_c\Delta V}\int_{\Delta V_c} d\bm{x}\int_{\Delta V} d\bm{c} \psi, \\
\Delta \psi&=&\psi-\overline{\psi},
\end{eqnarray}
where $\psi$ is the arbitrary function defined in $\zeta$-space.\\
On the basis of eqs. (5) and (6), we obtain the following equation by multiplying $\Delta \varrho(\beta)$ by both sides of eq. (4) and taking average in $\zeta$-space,
\begin{eqnarray}
&&\left[\frac{\partial}{\partial t(\alpha)}+\bm{c}(\alpha)\frac{\partial}{\partial \bm{x}(\alpha)}\right]\left[\overline{\varrho(\alpha)\varrho(\beta)}-f(\alpha)f(\beta)\right]\nonumber \\
&=&\mathscr{J}\left(\alpha|\hat{\alpha}\right)\left[\overline{\varrho(\alpha)\varrho(\hat{\alpha})\varrho(\beta)}-\overline{\varrho(\alpha)\varrho(\hat{\alpha})}f(\beta) \right],
\end{eqnarray}
where $f(\alpha)=\overline{\varrho(\alpha)} \wedge f(\beta)=\overline{\varrho(\beta)}$.\\
In eq. (7), $\overline{\varrho(\alpha)\varrho(\beta)}$ is decomposed as:
\begin{eqnarray}
\overline{\varrho(\alpha)\varrho(\beta)}=f_{II}(\alpha;\beta)+g(\alpha;\beta),
\end{eqnarray}
where $f_{II}\left(\alpha;\beta\right)$ is the two-point phase density of different particles and $g(\alpha;\beta)$ is the probability of finding same particles at $t=t(\alpha)\wedge \bm{\zeta}=\bm{\zeta}(\alpha)$ and $t=t(\beta) \wedge \bm{\zeta}=\bm{\zeta}(\beta)$. $f_{II}(\alpha;\beta)$ is related to the hydrodynamic fluctuation term $\phi(\alpha;\beta)$ by
\begin{eqnarray}
f_{II}(\alpha;\beta)&=&\phi(\alpha;\beta)+(1+\frac{1}{N})f(\alpha)f(\beta),\\
&\simeq& \phi(\alpha;\beta)+f(\alpha)f(\beta), (\mbox{For }1 \ll N),
\end{eqnarray}
when $1 \ll N$ and $t(\alpha) \rightarrow t(\beta)$, we obtain
\begin{eqnarray}
&&\lim_{t(\alpha) \rightarrow t(\beta)} f_{II}(\alpha;\beta)=\phi(\alpha,\beta)+f(\alpha)f(\beta),\\
&&\lim_{t(\alpha) \rightarrow t(\beta)} g(\alpha;\beta)=\delta\left[\xi(\alpha)-\xi(\beta)\right]f(\alpha)
\end{eqnarray}
Assuming that the hydrodynamic fluctuations is ignorable, (i.e., $\phi(\alpha;\beta)=0$ or $f_{II}(\alpha;\beta)=f(\alpha)f(\beta)$), we obtain the following equation from eq. (7):
\begin{eqnarray}
&&\left[\frac{\partial}{\partial \tau}+\bm{c}(\alpha)\frac{\partial}{\partial \bm{x}(\alpha)}\right]g\left(\alpha;\beta\right)=\mathscr{J}\left(\alpha|\hat{\alpha}\right)\left(f(\alpha)g(\hat{\alpha};\beta)+f(\hat{\alpha})g(\alpha;\beta)\right),
\end{eqnarray}
where $\tau \equiv t(\alpha)-t(\beta)$.\\
Tsuge-Sagara expanded $g(\alpha;\beta)$ using the Grad's method \cite{Grad} as follows:
\begin{eqnarray}
g(\alpha;\beta)=\omega(\alpha)\omega(\beta)\sum\frac{Q^{J,K}_{ij,...,lm...}}{\tilde{c}^{J+K}J!K!}H_{ij...}^{(J)}(\alpha)H_{lm...}^{(K)}(\beta)
\end{eqnarray}
where $\tilde{c}=\sqrt{RT}$, in which $R$ is the gas constant and $T$ is the temperature, $\omega(\alpha)=\frac{1}{(2\pi \tilde{c}^2)}^{\frac{3}{2}}\exp\left[-\frac{c(\alpha)^2}{2\tilde{c}^2} \right]$, and $H_{ij...}^{(J)}$ is the Hermite polynomial \cite{Grad}.
\begin{eqnarray}
Q_{ij...lm...}^{(J,K)}(\zeta(\alpha)-\zeta(\beta),\tau)=\int \tilde{c}^{J+K} H_{ij...}^{(J)}(\alpha)H_{lm...}^{(K)}(\beta)g(\alpha,\beta)dv(\alpha)dv(\beta).
\end{eqnarray}
We assume the initial equilibrium condition for $g(\alpha,\beta)$ using eq. (12) as follows
\begin{eqnarray}
[g(\alpha,\beta)]_{\tau=0}=\delta\left[\xi(\alpha)-\xi(\beta)\right]f_{MB}(\alpha),
\end{eqnarray}
where $f_{MB}$ is the Maxwell-Boltzmann distribution function.\\
Consequently, $Q$ is;
\begin{eqnarray}
Q_{ij...,lm...}^{(J,K)}(\bm{x}(\alpha)-\bm{x}(\beta),\tau)=n\tilde{c}^{J+K}\delta[\xi(\alpha)-\xi(\beta)]\delta_{JK}\sum \delta_{il}\delta_{jm}....
\end{eqnarray}
In particular, we have the correlation of the fluctuation of the number density and the correlation of the fluctuation of the velocity at $\tau=0$ as follows:
\begin{eqnarray}
[Q^{(0,0)}]_{\tau=0}&=&n \delta\left[\bm{x}(\alpha)-\bm{x}(\beta)\right]\nonumber \\
&&\mbox{(correlation of fluctuation of number density)},\\
\left[n^{-2}Q_{i,j}^{(1,1)}\right]_{\tau=0}&=&n^{-1}\tilde{c}^2\delta_{il}\delta\left[\bm{x}(\alpha)-\bm{x}(\beta)\right] \nonumber \\
&&\mbox{(correlation of fluctuation of velocity)}.
\end{eqnarray}
In a similar way, we have the correlation of the fluctuation of the static pressure, the correlation of the fluctuation of the pressure deviator, and the correlation of the fluctuation of the heat flux at $\tau=0$ as follows:
\begin{eqnarray}
\left[\overline{\Delta p(\alpha) \Delta p(\beta)}\right]_{\tau=0}&=&p^2 n^{-1}(5/3)\delta\left[\bm{x}(\alpha)-\bm{x}(\beta)\right] \nonumber \\
&&\mbox{(correlation of fluctuation of static pressure)} \\
\left[q_{ij,lm}^{(2,2)}\right]_{\tau=0}&=&\frac{p^2}{n}\delta\left[\bm{x}(\alpha)-\bm{x}(\beta)\right]\left(\delta_{il}\delta_{jm}+\delta_{jl}\delta_{im}-\frac{2}{3}\delta_{ij}\delta_{lm}\right)\nonumber\\
&&\mbox{(correlation of fluctuation of pressure deviator)}\\
\left[q_{i,j}^{(3,3)}\right]_{\tau=0}&=&\frac{5}{2}\frac{p^2}{n}RT \delta\left[\bm{x}(\alpha)-\bm{x}(\beta)\right] \delta_{il} \nonumber \\
&&\mbox{(correlation of fluctuation of heat flux)},
\end{eqnarray}
where $p=nkT$ is the static pressure.\\
13 moment equations \cite{Grad} are obtained by neglecting the convective terms from eq. (13) as follows:
\begin{eqnarray}
&&\frac{\partial}{\partial \tau}Q^{(0,0)}=\frac{\partial}{\partial \tau}Q_{i,l}^{(1,1)}=\frac{\partial}{\partial \tau}Q^{(2,2)}=0,\\
&&\frac{\partial}{\partial \tau}q_{ij,lm}^{(2,2)}+6nB q_{ij,lm}^{(2,2)}=0,\\
&&\frac{\partial}{\partial \tau}q_{i,l}^{(3,3)}+4nBq_{i,l}^{(3,3)}=0,
\end{eqnarray}
where $6nB=p/\mu$, in which $\mu$ is the viscosity coefficient for the elastic gas.\\
As a result, we obtain;
\begin{eqnarray}
q_{ij,lm}^{(2,2)}&=&n^{-1}p^2 \exp\left(-\tau p/\mu\right)\delta \left[\bm{x}(\alpha)-\bm{x}(\beta) \right]\left(\delta_{il}\delta_{jm}+\delta_{im}\delta_{jl}-\frac{2}{3}\delta_{ij}\delta_{lm}\right),\\
q_{i,l}^{(3,3)}&=&n^{-1}p^2 C_p T \exp\left(-\tau p C_p/\lambda\right)\left[\bm{x}(\alpha)-\bm{x}(\beta) \right] \delta_{il},
\end{eqnarray}
where $C_p=5R/2$ is the isobaric specific heat and $\lambda$ is the thermal conductivity of the elastic gas.\\
By approximating $e^{-a\tau} \sim (2/a)\delta(\tau)$ for $1 \ll a$, we obtain thermal fluctuations obtained by the FD theorem \cite{Landau}.
\section{Thermal fluctuations of granular gas using two-point kinetic theory}
The two-point kinetic theory of the elastic gas by Tsuge-Sagara was revisited. On the basis of forgoing discussions, we consider thermal fluctuations for the granular gas.\\
Different points from the elastic gas are;
\begin{enumerate}
\item The collision term defined in eqs. (1) and (2) is replaced by the inelastic Boltzmann equation.
\item The distribution function of the granular gas under the HCS is not the Maxwell-Boltzmann distribution.\\
Consequently, the initial self-correlation function $\left[g(\alpha,\beta)\right]_{\tau=0}=\delta(\xi(\alpha)-\xi(\beta))f^{(0)}$ is not expressed by the Maxwell-Boltzmann distribution as shown in eq. (16).
\end{enumerate}
To simplify our discussion, the restitution coefficient of the granular gas is constant in regardless of the relative velocity $g$. From the item "1", the collision term of the stochastic Boltzmann equation in eq. (2) is replaced by \cite{Brilliantov}:
\begin{eqnarray}
&&\mathscr{J}^{in}\left(\hat{\zeta}|\zeta\right)\left[\varrho\left(t,\zeta\right)\varrho\left(t,\hat{\zeta}\right)\right] \nonumber \\
&=&\int_{\mathscr{V}^3} d\hat{\bm{c}} \int_0^{2\pi} d\tilde{\epsilon} \int_0^\pi d\chi \left\{\frac{1}{\epsilon^2}\varrho\left(t,\zeta^{\prime\prime}\right)\varrho\left(t,\hat{\zeta}^{\prime\prime}\right)-\varrho\left(t,\zeta\right)\varrho\left(t,\hat{\zeta}\right) \right\}g\sigma \sin \chi, \nonumber \\
\end{eqnarray}
where the double prime indicates the inverse collision \cite{Brilliantov}, $\epsilon$ is the constant restitution coefficient and superscript $in$ indicates the inelastic collisions of the granular gas.\\
Under the HCS, we can start our discussion from eq. (13) by replacing the collision term $\mathscr{J}$ by $\mathscr{J}^{in}$ in eq. (28). As discussed in the item "2", the distribution function $f^{(0)}$ under the HCS does not obey the Maxwell-Boltzmann distribution function. Generally, $f^{(0)}$ under the HCS is expanded around the Maxwell-Boltzmann distribution as follows:
\begin{eqnarray}
f^{(0)}=f_{MB}\left(1+\frac{1}{8}a^{(4)}H^{(4)}+\frac{1}{5040}a^{(6)}H^{(6)}+\cdot\cdot\cdot +\theta^{2k}a^{2k}H^{(2k)}+\cdot\cdot\cdot \right),
\end{eqnarray}
where $a^{(2k)}$ is the spherically symmetric Grad's moment \cite{Yano} and $\theta^{(2k)}$ is the coefficient for a normalization. \\
For example, $H^{(4)}$ and $H^{(6)}$ are written as follows \cite{Yano}:
\begin{eqnarray}
H^{(4)}&=&v^4-10v^2+15,\\
H^{(6)}&=&v^6-21v^4+105v^2-105,
\end{eqnarray}
where $v^2=c^2/(RT)$.\\
On the basis of eq. (29), eqs (18)-(22) can be extended into the case of granular gas as follows:
\begin{eqnarray}
[Q^{(0,0)}]_{\tau=0}^{in}&=&n \delta\left[\bm{x}(\alpha)-\bm{x}(\beta)\right],\\
\left[n^{-2}{Q}_{i,j}^{(1,1)}\right]_{\tau=0}^{in}&=&n^{-1}\tilde{c}^2\delta_{il}\delta\left[\bm{x}(\alpha)-\bm{x}(\beta)\right],\\
\left[\overline{\Delta p(\alpha) \Delta p(\beta)}\right]_{\tau=0}^{in}&=&p^2 n^{-1}(5/3)\left(1+a^{(4)}\right)\delta\left[\bm{x}(\alpha)-\bm{x}(\beta)\right],\\
\left[q_{ij,lm}^{(2,2)}\right]_{\tau=0}^{in}&=&\frac{p^2}{n}\delta\left[\bm{x}(\alpha)-\bm{x}(\beta)\right]\left(1+a^{(4)}\right)\left(\delta_{il}\delta_{jm}+\delta_{jl}\delta_{im}-\frac{2}{3}\delta_{ij}\delta_{lm}\right),\\
\left[q_{i,l}^{(3,3)}\right]_{\tau=0}^{in}&=&\left(\frac{5}{2}+\frac{55}{4}a^{(4)}+\frac{1}{12}a^{(6)} \right)\frac{p^2}{n}RT \delta\left[\bm{x}(\alpha)-\bm{x}(\beta)\right] \delta_{il},
\end{eqnarray}
Moment equations for the elastic gas in eqs. (23)-(25) are extended into the granular gas as follows:
\begin{eqnarray}
&&\frac{\partial}{\partial \tau}\left[Q^{(0,0)}\right]^{in}=\frac{\partial}{\partial \tau}\left[Q_{i,l}^{(1,1)}\right]^{in}=0,\\
&&\frac{\partial}{\partial \tau}\left[Q^{(2,2)}\right]^{in}=-\tilde{\zeta} \left[Q^{(2,2)}\right]^{in},\\
&&\frac{\partial}{\partial \tau}\left[q_{ij,lm}^{(2,2)}\right]^{in}+6n\mathscr{S}^{(2,2)}(\epsilon)B \left[q_{ij,lm}^{(2,2)}\right]^{in}=0,\\
&&\frac{\partial}{\partial \tau}\left[q_{i,l}^{(3,3)}\right]^{in}+4n\mathscr{S}^{(3,3)}(\epsilon)B\left[q_{i,l}^{(3,3)}\right]^{in}=0,
\end{eqnarray}
where $\tilde{\zeta}$ is the cooling rate \cite{Brilliantov}.\\
$\mathscr{S}^{(2,2)}(\epsilon)$ and $\mathscr{S}^{(3,3)}(\epsilon)$ in eqs. (39) and (40) are given by Jenkins-Richman \cite{Jenkins} as follows:

\begin{eqnarray}
\mathscr{S}^{(2,2)}(\epsilon)&=&\frac{(1+\epsilon)(3-\epsilon)}{4},\\
\mathscr{S}^{(3,3)}(\epsilon)&=&\frac{(1+\epsilon)(49-33\epsilon)}{32}.
\end{eqnarray}
As a result of eqs. (35), (36), (39) and (40), we obtain
\begin{eqnarray}
\left[q_{ij,lm}^{(2,2)}\right]^{in}&=&\left(1+a^{(4)}\right)\frac{\exp\left(-\tau\mathscr{S}^{(2,2)}(\epsilon)p/\mu \right)}{\exp\left(-\tau p/\mu \right)}q_{ij,lm}^{(2,2)},\\
\left[q_{i,l}^{(3,3)}\right]^{in}&=&\left(1+\frac{11}{2}a^{(4)}+\frac{1}{30}a^{(6)} \right)\frac{\exp\left(-\tau\Pr\mathscr{S}^{(3,3)}(\epsilon)p/\mu \right)}{\exp\left(-\tau \Pr p/\mu \right)}q_{i,l}^{(3,3)},
\end{eqnarray}
where $\Pr=2/3$ is performed for the monatomic gas.\\
By approximating $e^{-a\tau} \sim (2/a)\delta(\tau)$ for $1 \ll a$, we obtain the deviations of thermal fluctuations of the granular gas from those for the elastic gas obtained by the FD theorem as follows:
\begin{eqnarray}
&&\left[q_{ij,lm}^{(2,2)}\right]^{in}=\underbrace{\frac{1}{\mathscr{S}^{(2,2)}(\epsilon)}\left(1+a^{(4)}\right)}_{S^{(2,2)}}q_{ij,lm}^{(2,2)},\\
&&\left[q_{i,l}^{(3,3)}\right]^{in}=\underbrace{\frac{1}{\mathscr{S}^{(3,3)}(\epsilon)}\left(1+\frac{11}{2}a^{(4)}+\frac{1}{30}a^{(6)}\right)}_{S^{(3,3)}}q_{i,l}^{(3,3)},
\end{eqnarray}
where $S^{(2,2)}$ is the function, which reveals the deviation of thermal fluctuations of the pressure deviator for the granular gas from those for the elastic gas, and $S^{(3,3)}$ is the function, which reveals the deviation of thermal fluctuations of the heat flux for the granular gas from those for the elastic gas. 
\section{Discussion}
Finally, $S^{(2,2)}$ and $S^{(3,3)}$ in eqs. (45) and (46) are considered in detail. $a^{(4)}$ and $a^{(6)}$ under the HCS is calculated by Brilliantov-P$\ddot{\mbox{o}}$schel as follows \cite{Brilliantov2}:
\begin{eqnarray}
&&a^{(4)}= \nonumber \\
&&\frac{-16(-1623+1934\epsilon+895\epsilon^2-364\epsilon^3+3510\epsilon^4-7424\epsilon^5+3321\epsilon^6-480\epsilon^7+240\epsilon^8)}{214357-172458\epsilon+112155\epsilon^2+25716\epsilon^3-4410\epsilon^4-84480\epsilon^5+34800\epsilon^6-5600\epsilon^7+2800\epsilon^8},\nonumber \\
&&a^{(6)}= \nonumber \\
&&\frac{13440(217-386\epsilon-669\epsilon^2+1548\epsilon^3+154\epsilon^4-1600\epsilon^5+816\epsilon^6-160\epsilon^7+80\epsilon^8)}{214357-172458\epsilon+112155\epsilon^2+25716\epsilon^3-4410\epsilon^4-84480\epsilon^5+34800\epsilon^6-5600\epsilon^7+2800\epsilon^8}, \nonumber \\
\end{eqnarray}
From eqs. (41), (42), (45), (46) and (47), $S^{(2,2)}$ and $S^{(3,3)}$ are plotted in Fig. 1.
\begin{center}
\includegraphics[width=.75\linewidth]{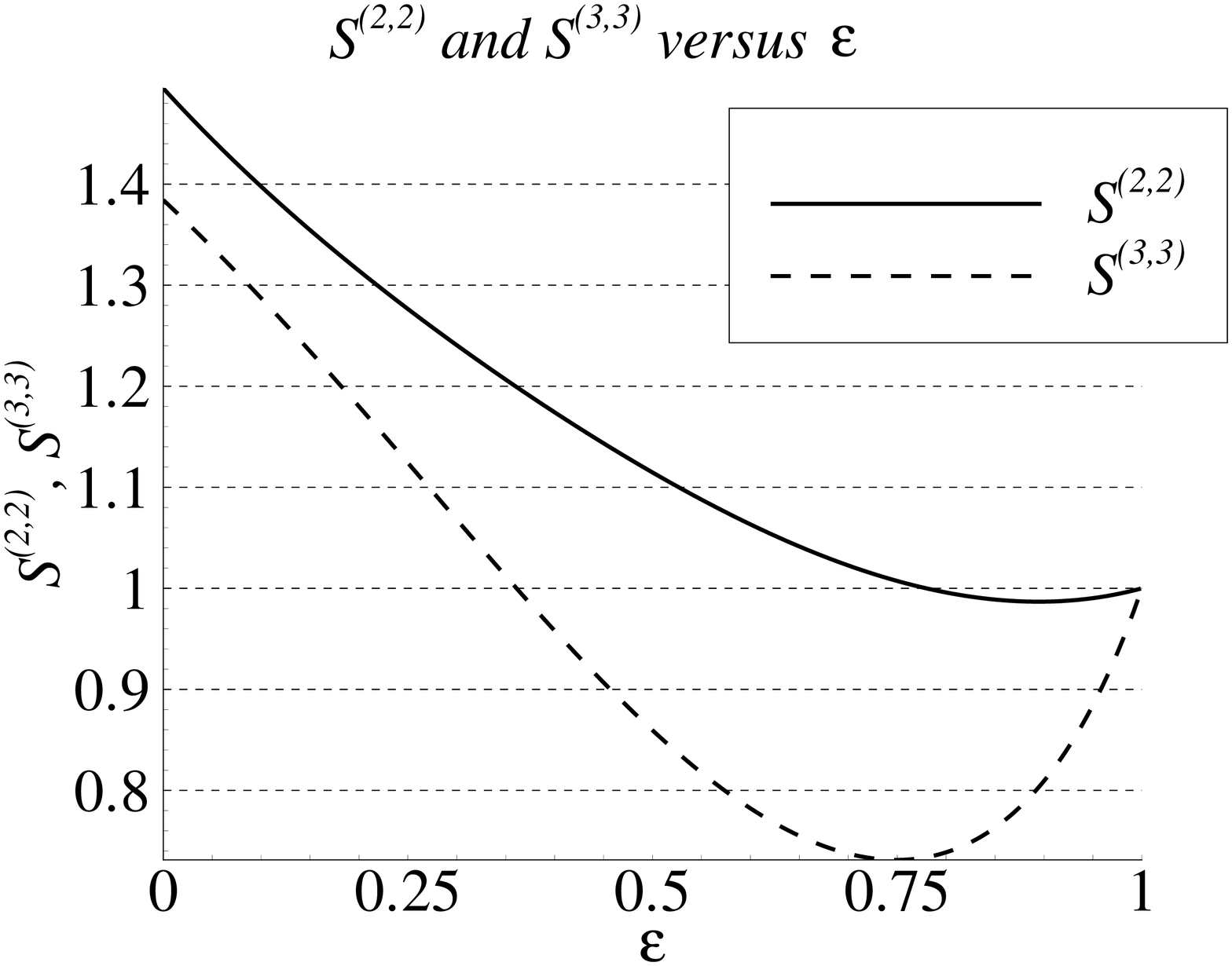}\\
\footnotesize{FIG. 1 $S^{(2,2)}$ and $S^{(3,3)}$ versus $\epsilon$ (restitution cofficient)}
\end{center}
$S^{(2,2)} \le 1$ at $0.786 \le \epsilon \le 1$ and $1 \le S^{(2,2)}$ at $0 \le \epsilon \le 0.786$ as shown in Fig. 1. $S^{(3,3)} \le 1$ at $0.36 \le \epsilon \le 1$ and $1 \le S^{(3,3)}$ at $0 \le \epsilon \le 0.36$ as shown in Fig. 1. $S^{(2,2)}$ has its minimum value $S^{(2,2)}_{\min}=0.986$ at $\epsilon=0.894$ and $S_{\min}^{(3,3)}$ has its minimum value $S_{\min}^{(3,3)}=0.731$ at $\epsilon=0.749$. The deviation-function $S^{(2,2)}$ does not change from unity around $\epsilon \sim 1$. As a result, the formulation of thermal fluctuations of the pressure deviator for the granular gas is obtained by slightly modifying that for the elastic gas in weakly inelastic regime (i.e., $\epsilon \sim 1$). On the other hand, the deviation-function $S^{(3,3)}$ drastically changes from unity around $\epsilon \sim 1$. As a result, the formulation of thermal fluctuations of the heat flux for the granular gas is obatined by remarakably modifying that for the elastic gas even in the weakly inelastic regime.

\end{document}